# Excitons in disordered polymers

I. Avgin[a] and D. L. Huber[b]

[a] Department of Electrical and Electronics Engineering,

Ege University, Bornova 35100, Izmir, Turkey

[b] Department of Physics, University of Wisconsin-Madison,

Madison, WI 53706 USA

## Abstract

We investigate the effects of disorder on Frenkel excitons in disordered conjugated polymers. In these materials, the principal effect of the disorder is to modify the transfer integrals appearing in the exciton Hamiltonian by changing the angle of relative rotation between polymeric units in a random manner. It is assumed the transfer integrals have the form $t\cos(\varphi_k - \varphi_{k+1})$ where $\varphi_k$ denotes the orientation angle of the k[th] unit. We focus on the two types of angular disorder: segmental disorder, which is characterized by infrequent, large fluctuations in $\Delta\phi$ and a continuous disorder marked by small fluctuations in each $\Delta\phi$. We calculate the density of states and inverse localization lengths for the exciton modes in long chains by mode-counting techniques and the absorption spectra in shorter chains by direct matrix diagonalization. Particular emphasis is placed on relating the features of the absorption spectra to the distribution and localization of the underlying exciton modes. Comparison is made with the spectra obtained assuming the disorder arises from random fluctuations in inter-unit bond length.

*Key words: polymers, excitons, absorption, photoluminescence, diso*rder

**D. L. Huber, Physics Dept., Univ. of Wisconsin-Madison, 1150 University Ave., Madison, WI 53706, e-mail huber@src.wisc.edu, phone (608) 265-4035, fax (608) 265-2334**

## I. Introduction

The possibility of using polymers in various photonic applications such as flat panel displays and LEDs has led to increased interest in their electronic and optical properties. Among the important classes of optically active polymers are conjugated systems with alternating single and double carbon-carbon bonds such as the polyfluorenes. Due to the nature of the bonding, which is analogous to the alternating bonding in benzene, the polymeric units have a planar or "platelet" character. Each platelet is connected to neighboring identical platelets by a single bond. Rotation about the bond can take place which leads to a situation where the individual platelets are characterized by local normal directions. In situations where the electron-phonon interactions are weak, the low-lying optically active states are frequently singlet excitons involving the $\pi$ and $\pi^*$ orbitals of the platelets. In this paper we report the results of theoretical studies of the effects of disorder on the distribution and localization of the excitonic states and the corresponding excitonic spectra.

Our approach is based on the Frenkel exciton picture in which we regard each polymeric unit as a single chromaphore, analogous to the situation in molecular crystals. The corresponding Hamiltonian takes the form

$$H = \sum_k E_k c_k^* c_k - \sum_k t_{k,k+1}(c_{k+1}^* c_k + c_k^* c_{k+1}) \qquad (1)$$

where $E_k$ denotes the chromaphore transition energy, $t_{k,k+1}$ is the transfer integral, and the $c_k$ and $c_k^*$ are exciton operators. In our analysis, we assume all of the transition energies are identical (i.e. no diagonal disorder) and that the transfer integral is proportional to the cosine of the angle between the normal vectors of platelets $k$ and $k+1$ [1]. That is we have

$$t_{k,k+1} = t\cos(\varphi_{k+1}-\varphi_k) \equiv t\cos(\Delta\varphi_{k,k+1}) \qquad (2)$$

where $\Delta\varphi_{k,k+1}$ is the relative angle of rotation between platelets $k$ and $k+1$ [1]. The corresponding Hamiltonian has the form

$$H = \sum_k E_0 c_k^* c_k - \sum_k t\cos(\Delta\varphi_{k,k+1})(c_{k+1}^* c_k + c_k^* c_{k+1}) \qquad (3)$$

In this model, the disorder arises from random fluctuations in $\Delta\varphi$. Two types of rotational disorder are considered: segmental disorder where $\Delta\varphi = 0$, except for large, *random* excursions [1], and continuous modulation in which there are small random fluctuations in $\Delta\varphi$ between adjacent platelets. In the analysis we compare our results obtained assuming angular disorder with the corresponding results arising from random variations in bond length with no angular disorder.

In the standard treatment of Frenkel excitons, the Hamiltonian (3) is transformed into a set of coupled linear equations that take the form

$$(E - E_k)a_k = -t\cos(\Delta\varphi_{k,k+1})a_{k+1} - t\cos(\Delta\varphi_{k-1,k})a_{k-1}, \qquad (4)$$

with eigenvalues $E^\nu$ and orthonormal eigenvectors $a_k^\nu$, $\nu = 1, 2,\ldots, N$, for a chain of N platelets.

The distribution and localization of the exciton modes are characterized by the density of states and the inverse localization length, respectively, with the latter being the reciprocal of the number of platelets where the mode has significant amplitude. In the case of one-dimensional arrays with nearest-neighbor interactions, both of these quantities can be determined by making use of mode-counting techniques [2,3]. In detail, one introduces the amplitude ratios $R_k = a_k/a_{k-1}$ and calculates $R_k$ from the recursion relation

$$(E - E_k) = -t\cos(\Delta\varphi_{k,k+1})R_{k+1}(E) - t\cos(\Delta\varphi_{k-1,k})/R_k(E) \qquad (5)$$

The density of states (DOS) and the inverse localization length (ILL) are expressed in terms of the quantity

$$\gamma(E) = (1/N)\Sigma_n^N \ln[R_n(E)] \tag{6}$$

by means of the equations [4]

$$ILL(E) = Re\,\gamma(E) \tag{7a}$$

$$DOS(E) = (1/\pi)d[Im\,\gamma(E)]/dE \tag{7b}$$

where $Re$ and $Im$ denote real and imaginary parts, respectively.

The absorption spectra are calculated from the equation

$$I(E) = |\mu|^2 \Sigma_{\nu=1}^N |\Sigma_k a_k^\nu|^2 \delta(E - E_\nu) \tag{8}$$

where $\mu$ is the optical transition matrix element. Note that unlike the DOS and ILL, the absorption spectra is calculated by diagonalizing the dynamical matrix associated with Eq. (4). In practice, one replaces the delta function appearing in Eq. (8) with a Lorentzian or similar broadening function.

## II. Model Calculations

In the model calculations, we assume that the angular differences between adjacent platelets, $\Delta\varphi$, are distributed at random according to the equation

$$Prob(\Delta\varphi_{k+1,k} = \Delta\varphi) = c, \quad Prob(\Delta\varphi_{k+1,k} = 0) = 1 - c \tag{9}$$

with no correlation between sites. We note that when $\Delta\varphi = \pi/2$ ($\cos\Delta\varphi = 0$), the effect of the fluctuations is to divided the polymer into ideal, independent segments of average length $1/c$, a situation that has been studied in detail by various authors (*e.g.* [5]). The case $\Delta\varphi = \pi$ is unusual in that a mathematical analysis, confirmed by numerical studies, shows that the density of states is the same as in an ideal chain ($c = 0$), and all of the modes are delocalized (ILL = 0). The eigenvectors, however, are not the same as for the ideal chain, leading to optical spectra that broaden with increasing $c$ (see Fig. 1).

In Figs. 2 – 4 we display the density of states, the inverse localization length and the optical spectra for $\Delta\varphi = \pi/4$ with $c = 0$, 0.1, 0.25 and 0.50. We see that the effects of disorder on the DOS and ILL show up most strongly at the band edges. Toward the center the states are more extended and the distribution resembles that of the ideal chain. The optical absorption is strongest near the lower band edge, consistent with the direct absorption. With increasing disorder, the peak broadens and shifts toward the center of the band. Although the spectra were calculated with one value of $\Delta\varphi$, the results are representative of the results obtained for other large, discrete changes in relative platelet orientation apart from $\Delta\varphi = \pi/2$ and $\pi$.

As noted above, we have also explored the effects of small fluctuations in the relative orientations of nearest-neighbor platelets. This was accomplished by giving $\Delta\varphi$ a Gaussian distribution centered at 0º with no correlation between sites. The results for the absorption spectra with variance $\sigma^2$, $\sigma = 0º$, 6º and 15º, are shown in Fig. 5. From this figure it is evident that the effects of continuous disorder resemble those of segmental disorder in that the peak in absorption broadens and moves toward the center of the band with increasing disorder.

At high temperatures, one may approach a situation where $\Delta\varphi$ is uniformly distributed between 0 and $2\pi$. In this limit the excitons are strongly localized (except at the center of the band [6]), and the absorption spectrum resembles the density of states, which is shown in Fig. 6. One can understand this behavior from Eq. (8) by expanding $|\Sigma_k a_k^\nu|^2$. In a strongly disordered system, we have random phases so that terms in the expansion of the form $a_k^\nu \, a_{k'}^{\nu}{}^*$ average to zero for $k \neq k'$. Thus, the expression $|\Sigma_k \, a_k^\nu|^2$ reduces to $\Sigma_k \, |a_k^\nu|^2 = 1$, since the eigenvectors are normalized. In this limit we obtain

$$I(E) = |\mu|^2 \, \Sigma_{\nu=1}^{N} \delta(E - E_\nu) \equiv |\mu|^2 \text{DOS}(E) \qquad (10)$$

In other words, I(E) is proportional to the density of states.

## III. Discussion

The most direct information about the excitons in conjugated polymers comes from studies that probe the absorption lineshape. As noted, isolated, large fluctuations and continuous small fluctuations in $\Delta\varphi$ have qualitatively similar effects on the lineshape. In both cases the effect of the disorder is to broaden and shift the peak toward the center of the band. This behavior is in contrast to the behavior found when the disorder arises from fluctuations in bond length. This can be seen in a model calculation where $\Delta\varphi = 0$ and the transfer integral varies as

$$t = 1 + X \qquad (11)$$

where X has a Gaussian distribution with zero mean and variance $\sigma^2$. The absorption lineshapes and the corresponding densities of states calculated with $\sigma = 0, 0.10, 0.25$ and 0.50 are shown in Figs. 7 and 8, respectively. It is apparent there are qualitative differences between the bond angle disorder and bond length disorder. In the latter, the absorption peak shifts away from the band center with increasing disorder and develops exponential tails. The difference between Fig. 7 and Figs. 4 and 5 can be traced to the property that the magnitude of the transfer integral is bounded in the latter two cases, $-1 \leq t_{k+1,k} \leq 1$, so that all of the eigenvalues lie between $-2$ and $2$ as in systems with no disorder. With the Gaussian distribution of bond lengths, the eigenvalues are not bounded, and thus the states tail off from the upper and lower band edges (with a singularity at the center of the band [6]). Thus with angular disorder, one predicts a well defined onset for the absorption whereas in the case of bond length disorder, one has a lineshape that reflects the tailing of the density of states. Note that a similar lineshape is also obtained with a Gaussian distribution of transition energies, *i.e.* diagonal disorder [7].

Evidence that supports the picture of angular disorder comes from measurements of the absorption and photoluminescence of the polyfluorene PF8 [8,9]. In these references, it was pointed out that the photoluminescence at low temperatures comes from a thermally populated band of β-exciton states associated with an absorption lineshape that varies as $E^{3/2}$ near the edge. The authors of Ref. 9 were able to model the emission and photoluminescence profiles in an approach that incorporates both segmental disorder (due to the admixture of α and β phases with different torsion angles) and continuous disorder.

## Acknowledgment

DLH would like to thank M. Winokur for many helpful comments and insights into the physics of polymers.

# Figure Captions

Fig. 1. Absorption intensity vs $E$ for $\Delta\varphi = \pi$ with $c = 0$ (no disorder), 0.10, 0.25 and 0.50 (Eq. (9)). Note that the spectral weight shifts toward the center of the band with increasing disorder.

Fig. 2. Density of states vs $E$ for $\Delta\varphi = \pi/4$ with $c = 0$, 0.10, 0.25 and 0.50 (Eq. (9)). Note that the spectral weight shifts toward the center of the band with increasing disorder, and there are no states with $|E| > 2$.

Fig. 3. Inverse localization length vs $E$ for $\Delta\varphi = \pi/4$ with $c = 0.10$, 0.25 and 0.50 (Eq. (9)). Note that the excitons are most localized near the band edges and the ILL increases with increasing disorder.

Fig. 4. Absorption intensity vs $E$ for $\Delta\varphi = \pi/4$ with $c = 0$, 0.10, 0.25 and 0.50 (Eq. (9)). Note that the peak shifts toward the center of the band with increasing disorder.

Fig. 5. Absorption intensity vs $E$ for a Gaussian distribution of the fluctuations in $\Delta\varphi$. Results are presented for different various values of the variance $\sigma^2$: $\sigma = 0°$ (no disorder), $6°$ and $15°$. Note that the peak shifts toward the center of the band with increasing disorder.

Fig. 6. Density of states vs $E$ for a uniform distribution of the fluctuations in $\Delta\varphi$, $0 \leq \Delta\varphi \leq 2\pi$. Note there are no states with $|E| > 2$.

Fig. 7. Absorption intensity vs $E$ for bond length disorder. Results are presented for various values of the variance $\sigma^2$: $\sigma = 0$ (no disorder), 0.1, 0.25 and 0.50 (Eq. (11)). Note that the peak shifts away from the center of the band with increasing disorder in contrast to what happens in the case of angular disorder.

Fig. 8. Density of states vs $E$ for bond length disorder. Results are presented for various values of the variance $\sigma^2$: $\sigma = 0$, 0.1, 0.25 and 0.50 (Eq. (11)). Note the tailing into the regions $|E| > 2$ with increasing disorder

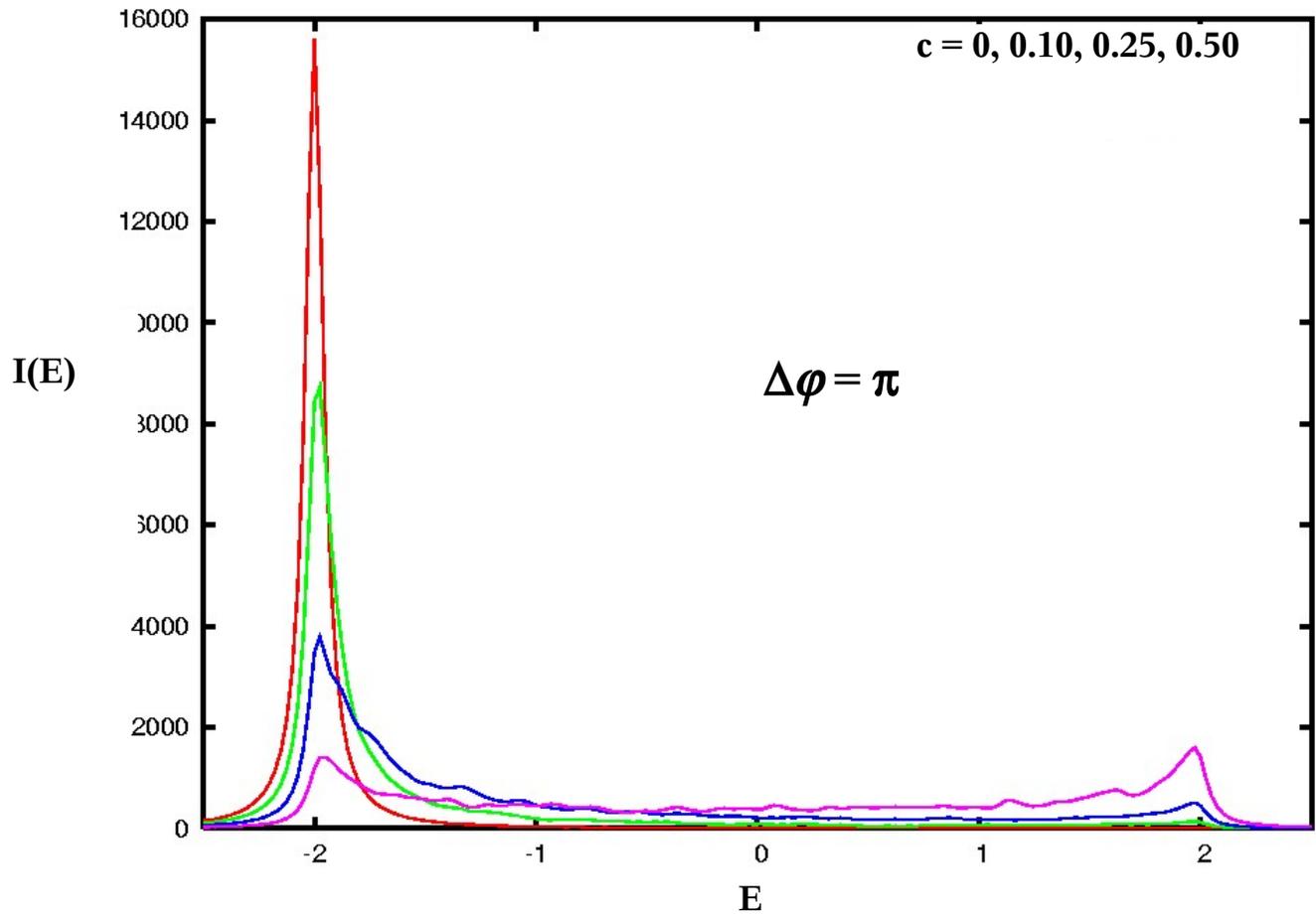

Fig. 1

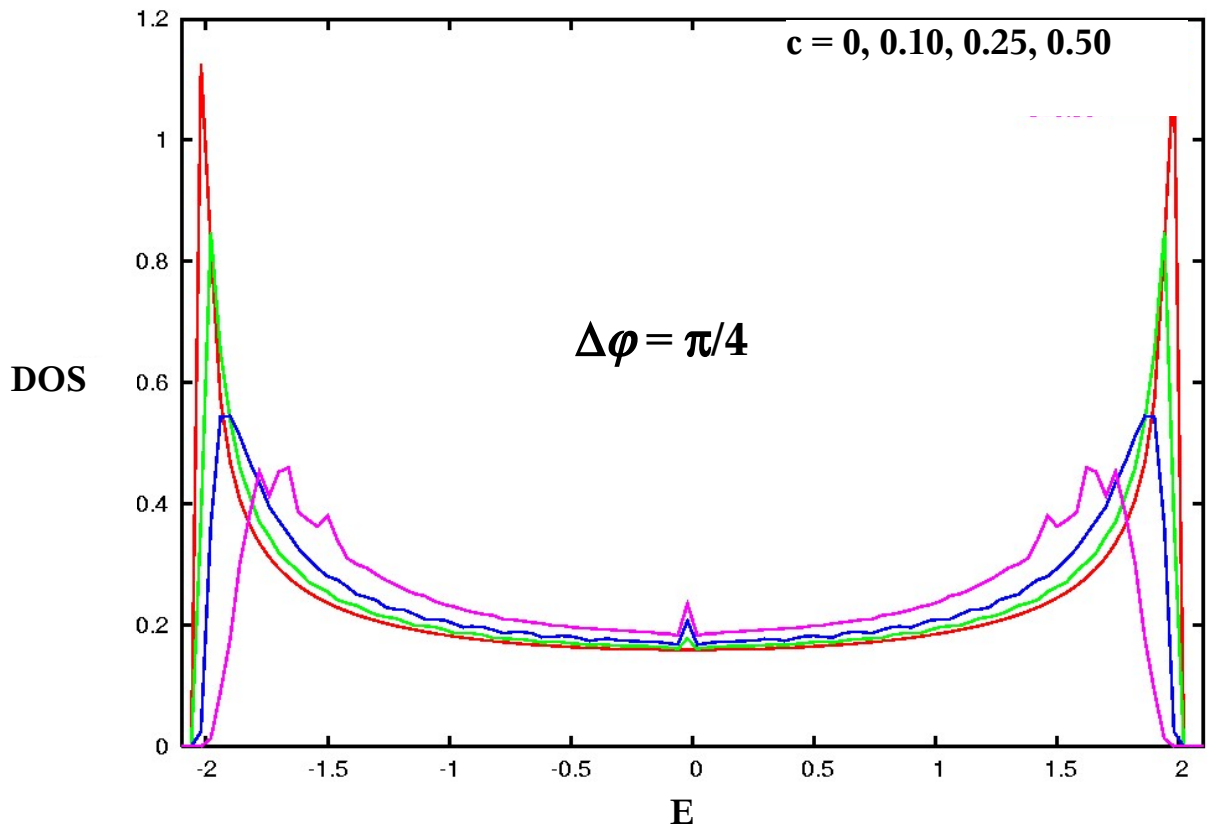

Fig. 2

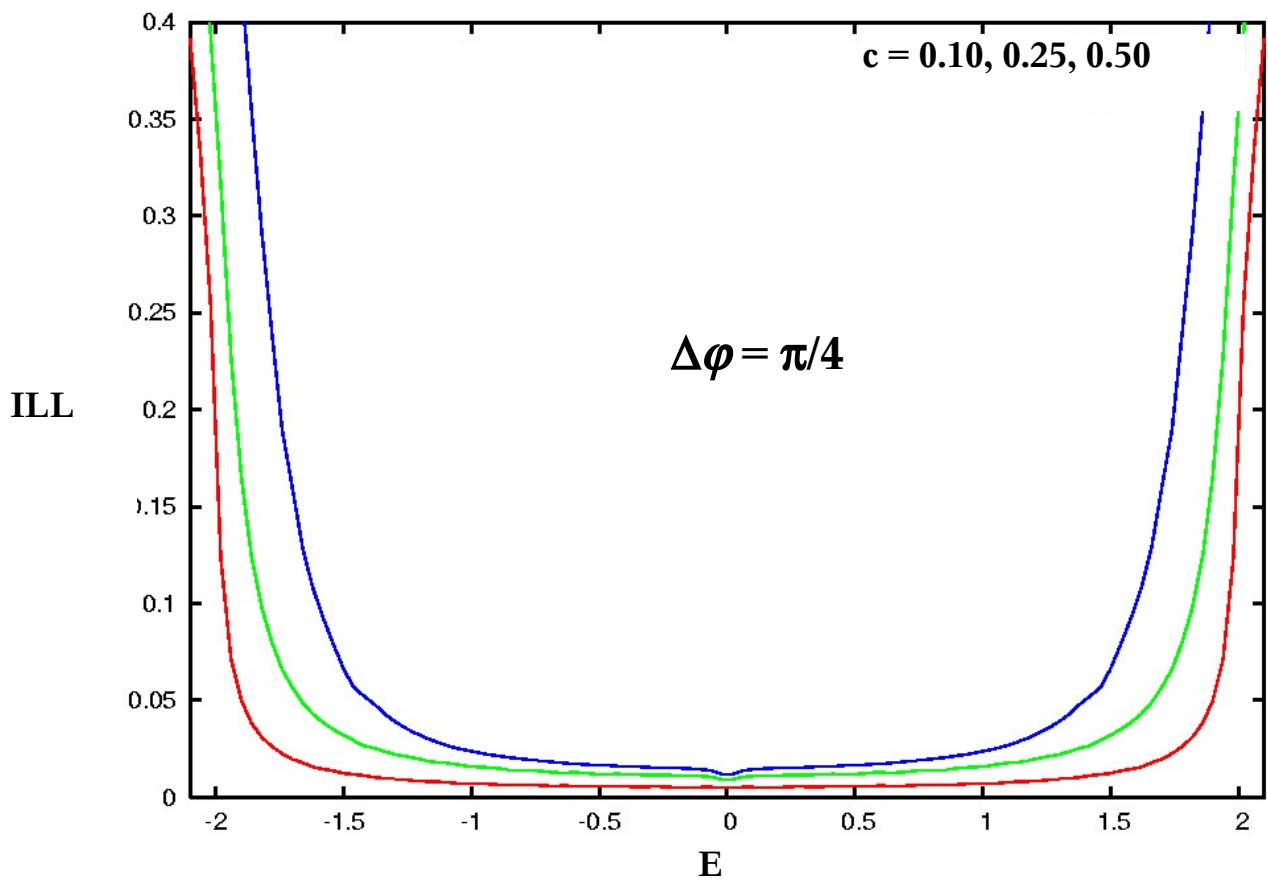

Fig. 3

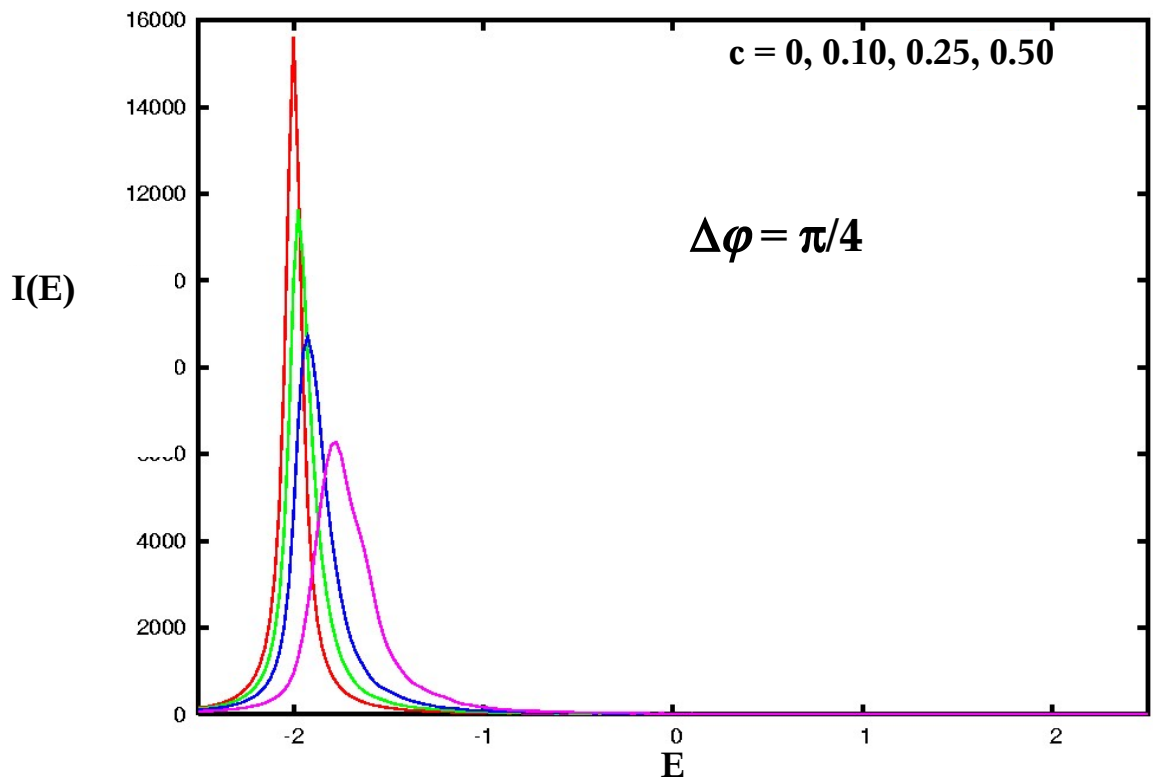

Fig. 4

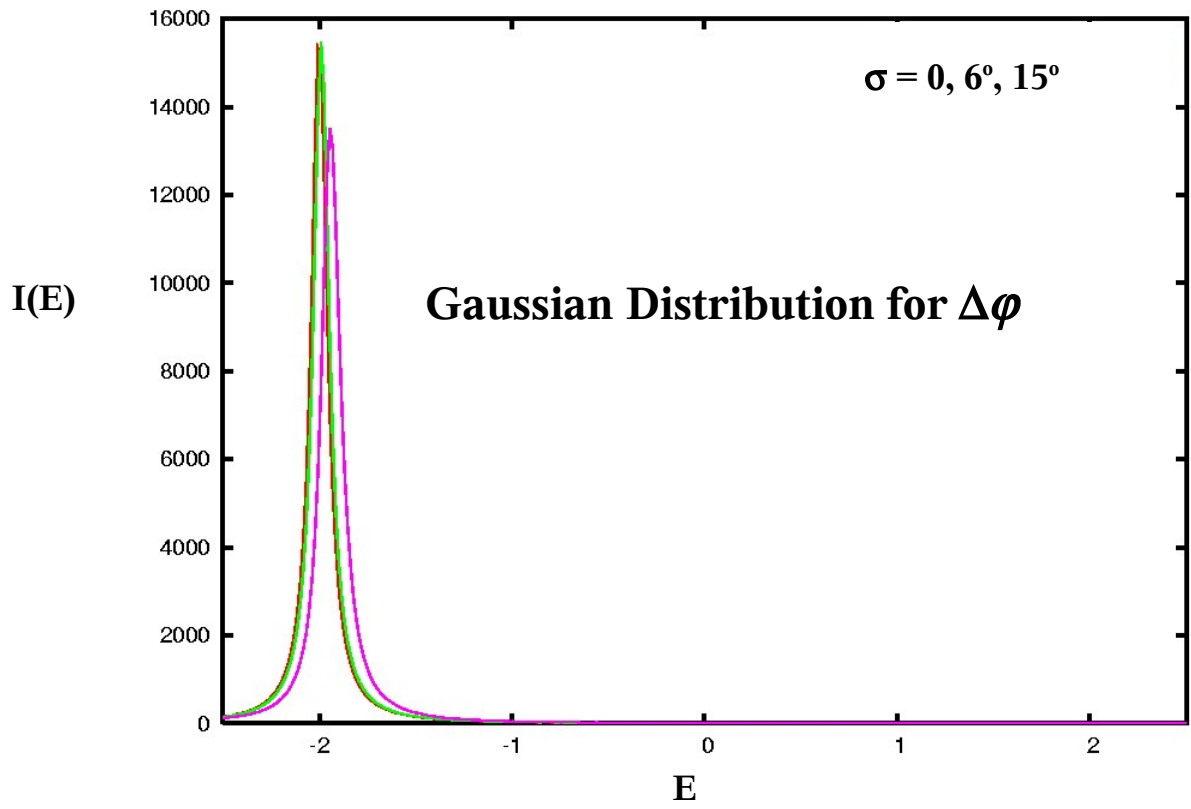

Fig. 5

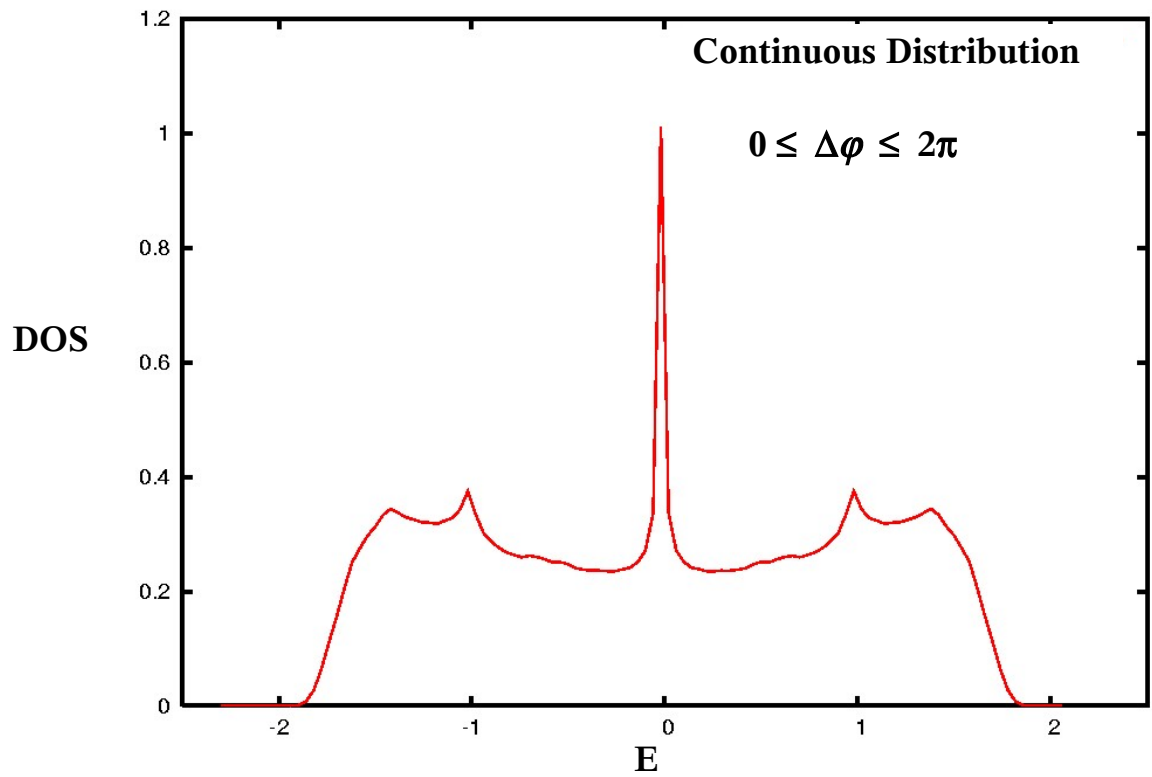

Fig. 6

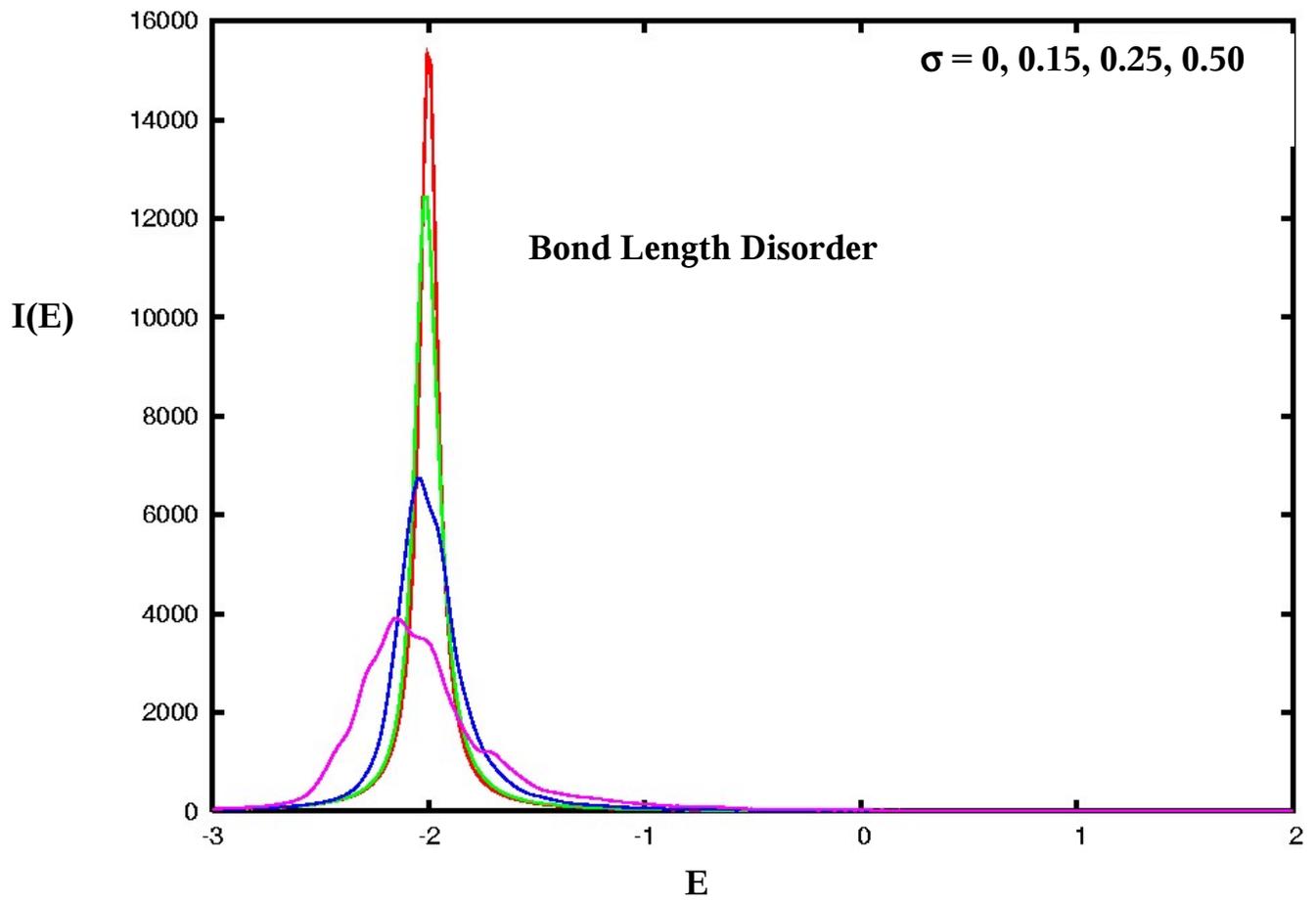

Fig. 7

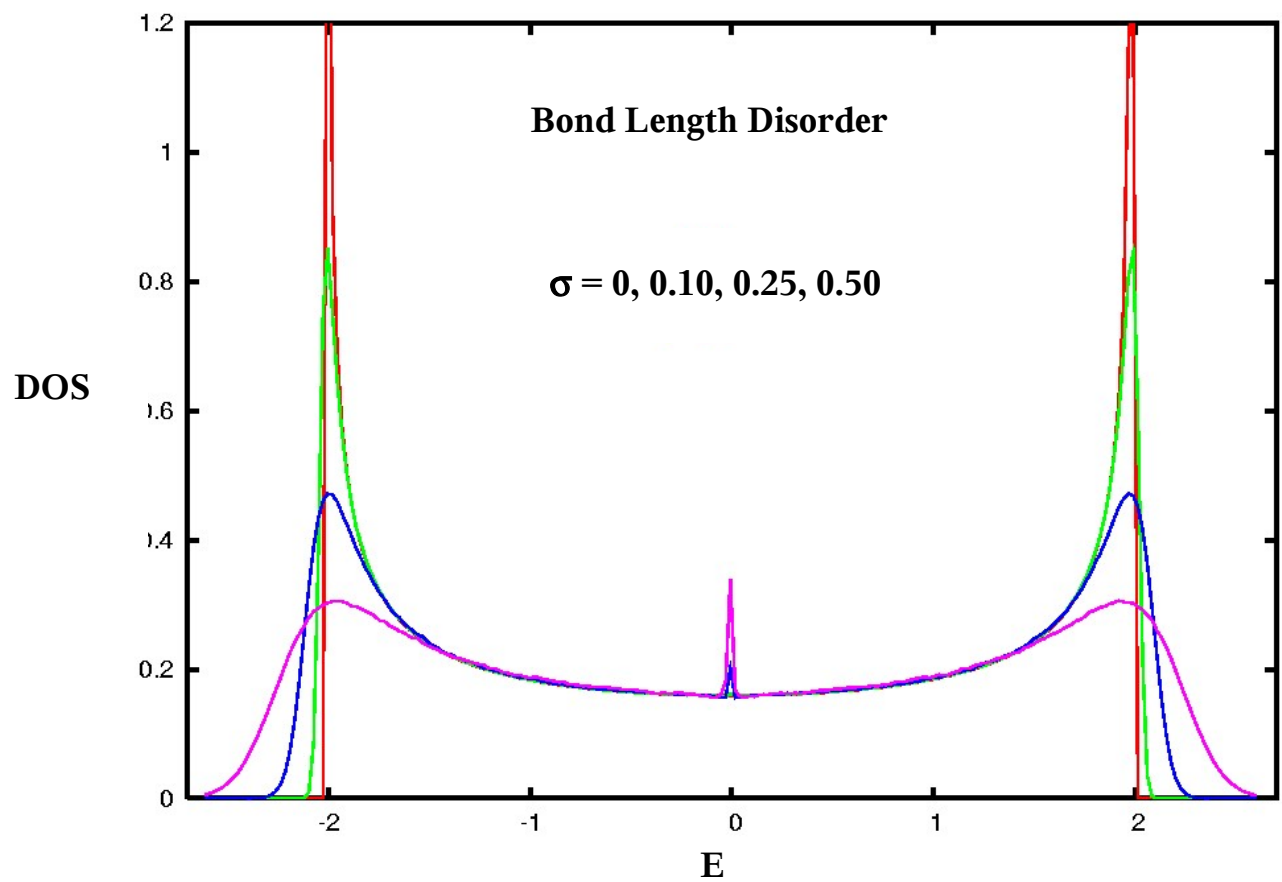

Fig. 8